\documentclass[12pt]{iopart}
\usepackage{color}
\usepackage{bm}
\usepackage{caption}
\usepackage[noadjust]{cite}
\usepackage{graphicx}

\begin{document}

\title{What do we learn about vector interactions from GW170817?}

\author{Veronica Dexheimer}
\address{Department of Physics, Kent State University, Kent OH 44242 USA}
\ead{vdexheim@kent.edu}

\author{Rosana de Oliveira Gomes, Stefan Schramm}
\address{Frankfurt Institute for Advanced Studies, Frankfurt am Main, Germany}

\author{Helena Pais}
\address{CFisUC, Department of Physics, University of Coimbra, 3004-516 Coimbra, Portugal}

%\vspace{10pt}
%\begin{indented}
%\item[]August 2017
%\end{indented}

\begin{abstract}
We analyze the role played by a mixed vector-isoscalar/vector-isovector meson interaction in dense matter present in the interior of neutron stars in the light of new measurements made during the double neutron-star merger GW170817. These concern measurements of tidal deformability from gravitational waves and electromagnetic observations. Our study includes three different equations of state that contain different physical assumptions and matter compositions, namely the NL3 family, MBF, and CMF models. Other related quantities/relations analyzed are the neutron matter pressure, symmetry energy slope, stellar masses and radii, and Urca process threshold for stellar cooling. 
\end{abstract}

\vspace{2pc}
\noindent{\it Keywords}: Neutron Star; Vector Interaction; Tidal Deformability; Gravitational Waves.
%\maketitle

\section{Introduction}

Since the first accurate measurement of a large-mass neutron star \cite{Demorest:2010bx}, extensive discussion in the nuclear astrophysics community has been devoted to the strength and kind of interactions among the nucleons and hyperons in dense matter. These interactions were, for example, presented as a solution for the so-called hyperon puzzle \cite{Buballa:2014jta}. A few years later, the discussion resurfaced based on analyses of symmetry energy slope values \cite{Lattimer:2012xj,Li:2013ola,Horowitz:2014bja} that pointed to neutron star radii smaller than previously assumed, together with more accurate estimates of neutron star radii \cite{Psaltis:2013fha,Guillot:2013wu,Lattimer:2013hma,Heinke:2014xaa}. Note, however, that other works such as the ones in Ref.~\cite{Sotani:2015lya,Cao:2015lda,Danielewicz:2016bgb} and references therein found larger values for the symmetry energy slope.

More recently, after the first detection of gravitational waves from a double neutron-star binary \cite{TheLIGOScientific:2017qsa,Monitor:2017mdv}, new quantities related to compact star mergers were calculated. For microscopic modeling, the most important of these is the tidal deformability ${\Lambda}$. The understanding of tidal deformation in double neutron stars systems has been a topic in the literature for some time \cite{Hinderer:2009ca, Damour:2009vw, Postnikov:2010yn}. Recently, extensive research has been performed in order to understand the composition of neutron stars \cite{Kumar:2016dks,Drago:2017bnf,Nandi:2017rhy,Alford:2017qgh,Marques:2017zju,Banik:2017zia,Ayriyan:2017nby,Paschalidis:2017qmb,Most:2018hfd,Burgio:2018yix,Alvarez-Castillo:2018pve,Rueda:2018fky,Gomes:2018eiv} 
and also on constraining the nuclear matter equation of state (EoS) 
\cite{Banik:2017zia,Alford:2017qgh,Annala:2017llu,Zhou:2017pha,Drago:2018nzf,Krastev:2018nwr,Zhu:2018ona,Raithel:2018ncd,Malik:2018zcf,Sieniawska:2018zzj,Sun:2018tmw,Zhao:2018nyf,Li:2018ayl}. The value of the dimensionless tidal deformability $\tilde{\Lambda}$ has recently been calculated to be in the range 76-1045 with $90\%$ confidence, under different assumptions \cite{De:2018uhw}. This corresponds to a radius between $8.7$ and $14.1$ km for a $1.4$ M$_\odot$ mass star. Other references found different values, as for example a lower bound for the radius $\sim10$ km in Ref.~\cite{Bauswein:2017vtn}. Note that assumptions of strong first-order phase transitions, for example, can modify these ranges \cite{Most:2018hfd}. Electromagnetic observations can help to provide, on the other hand, lower bounds for the tidal deformability \cite{Radice:2017lry}.

In this work, we study the role played by the $\omega\rho$ vector interaction and its implications for the microscopic and macroscopic properties of neutron stars focusing on new results related to GW170817. In the past, other works have already pointed to the importance of such an interaction and its relation to neutron stars \cite{Schramm2003,Dexheimer:2008ax,Bednarek:2011,Steiner:2012rk,Logoteta:2013ipa,Dutra:2014qga,Bizarro:2015wxa,Dexheimer:2015qha,Zhao:2015ncr,Pais:2016xiu,Tolos:2016hhl,Zhao:2017pim,Hornick:2018kfi}, although it was first introduced in the context of the
thickness of neutron skin in $^{208}Pb$ \cite{Horowitz2002}. It has been recently shown that crossed scalar terms of the kind $\sigma\delta$ in the Lagrangian density also affect the symmetry energy slope \cite{Zabari:2018tjk}.

\section{Equations of State}

In the following, we briefly describe the three EOS models used in our study. They are all in agreement with standard constraints from nuclear physics and astrophysics. The three models include the following Lagrangian term for the coupling of the vector mesons $L_{\omega\rho}=g_{\omega\rho} g_\omega^2 g_\rho^2 \omega_\mu \omega^\mu {\bm{\rho_\mu \rho^\mu}}$. Here, $g_\omega$ and $g_\rho$ are the coupling constants of the vector-isoscalar and vector-isovector mesons, respectively.

\subsection{NL3 family}

NL3 \cite{NL3} is a relativistic mean field model based on the nuclear Walecka model, in which the nucleons interact with the isoscalar-scalar $\sigma$, isoscalar-vector $\omega$ and isovector-vector $\rho$ mesons. The parameters of the Lagrangian density were fitted to reproduce the properties of 10 stable nuclei, total binding energies, charge radii, and neutron radii. As a result, the slope of the symmetry energy at saturation is high, 118 MeV.  The NL3 family \cite{Carriere:2002bx,Pais:2016xiu} was then constructed in order to model the density-dependence of the symmetry energy, making the EoS softer by decreasing the slope of the symmetry energy. This was done by adding a non-linear term that couples together the $\omega$ and $\rho$ mesons. This family of models is able to describe  two-solar-mass stars and it also fulfills constrains coming from state-of-the-art microscopic neutron matter calculations \cite{Hebeler,Gandolfi}.

\subsection{MBF}

The many body forces (MBF) is a relativistic mean-field model that includes a scalar field dependence in the coupling of baryons to mesons. The field dependence is introduced as a medium effect, which also makes the model indirectly density dependent. The octet of baryons interacts through the scalar $\sigma$ and (isovector) $\delta$ mesons and the vector $\omega$, $\rho$ and (isoscalar) $\phi$ mesons, with the $\phi$ meson (anti-strange/strange quark state said to have hidden strangeness) interacting only with the hyperons. By $\delta$ meson, we are referring to the $a_0(980)$ state, while the relation between $\sigma$ and the $f_0(500)$ is more loose. In this work, we use a stiff parametrization of the model $\zeta=0.040$ that reproduces  nuclear-matter properties (such as density, binding energy, compressibility, and symmetry energy), and also describes large-mass hyperonic stars \cite{Gomes:2014aka}. In this work, we include the $\omega\rho$ self-interaction in the MBF model for the first time.

\subsection{CMF}

The Chiral Mean Field (CMF) model is based on a non-linear realization of the SU(3) sigma model. It is an effective quantum relativistic model that describes hadrons and quarks interacting via meson exchange and it is constructed in a chirally-invariant manner, as the particle masses originate from interactions with the medium and, therefore, decrease at high densities/temperatures \cite{Papazoglou:1998vr,Dexheimer:2008ax}.  The hadronic coupling constants of the model were calibrated to reproduce the vacuum masses of baryons and mesons, and were fitted to reproduce nuclear constraints for isospin symmetric matter and symmetry energy  at saturation together with reasonable values for the hyperon potentials and two solar mass stars. The quark coupling constants were constrained using lattice QCD data, as well as information about the QCD phase diagram for isospin symmetric matter and tested against perturbative QCD \cite{Dexheimer:2009hi,Hempel:2013tfa,Kurkela:2016was,Roark:2018uls}. As a consequence, this formalism reproduces the nuclear liquid-gas phase transition and exhibits deconfinement/chiral symmetry restoration phase transitions expected to be found in the QCD phase diagram. In this work, we focus on the hadronic version of the model, as it is the one pertinent for the study of low-mass stars.

The CMF formalism models the interaction among the hadrons through a mean field of mesons, the scalars $\sigma$, $\delta$, and (isoscalar with hidden strangeness) $\zeta$, and the vectors $\omega$, $\rho$, and, $\phi$. By $\zeta$ meson, we are referring to the $f_0(980)$ state, also referred to as  $\sigma^*$ meson. Specially interesting for stellar mass and radius comparisons are the mesonic self-vector interactions, which have already been briefly studied in Ref.~\cite{Dexheimer:2008ax}. Additionally, in Ref.~\cite{Dexheimer:2015qha}, we have looked at different options for these couplings, which allowed for a lower symmetry energy slope. Here, we do something different and add separately a $\omega\rho$ self-interaction to the existing standard CMF vector-self-coupling terms $\omega$, $\phi$, and $\omega\phi$.

\begin{figure}[t!]
\centering
%\begin{subfigure}{0.495\textwidth}
%\centering
\includegraphics[width=0.496\textwidth]{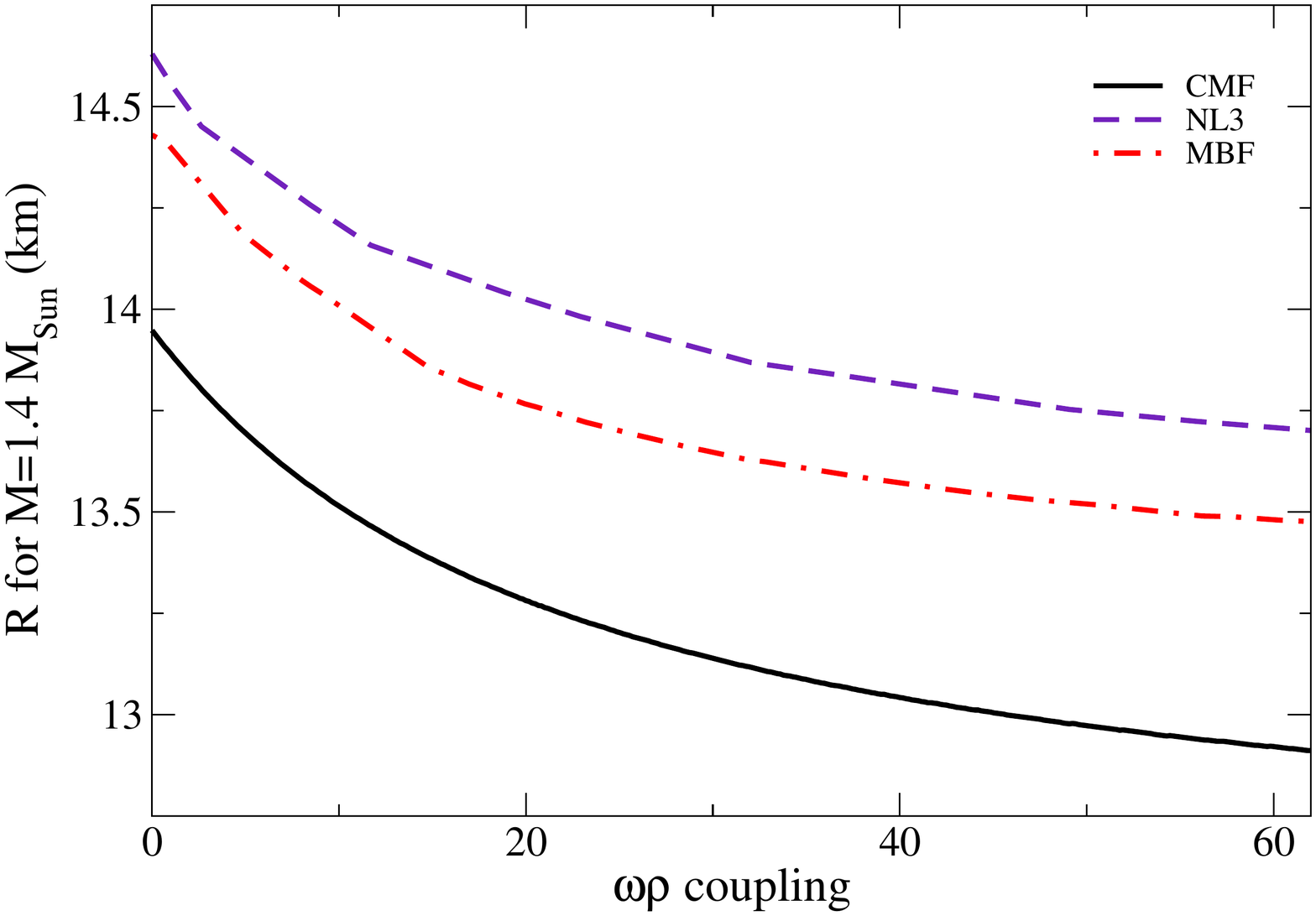} 
%\end{subfigure}
%
%\begin{subfigure}{0.495\textwidth}
%\centering
\includegraphics[width=0.496\textwidth]{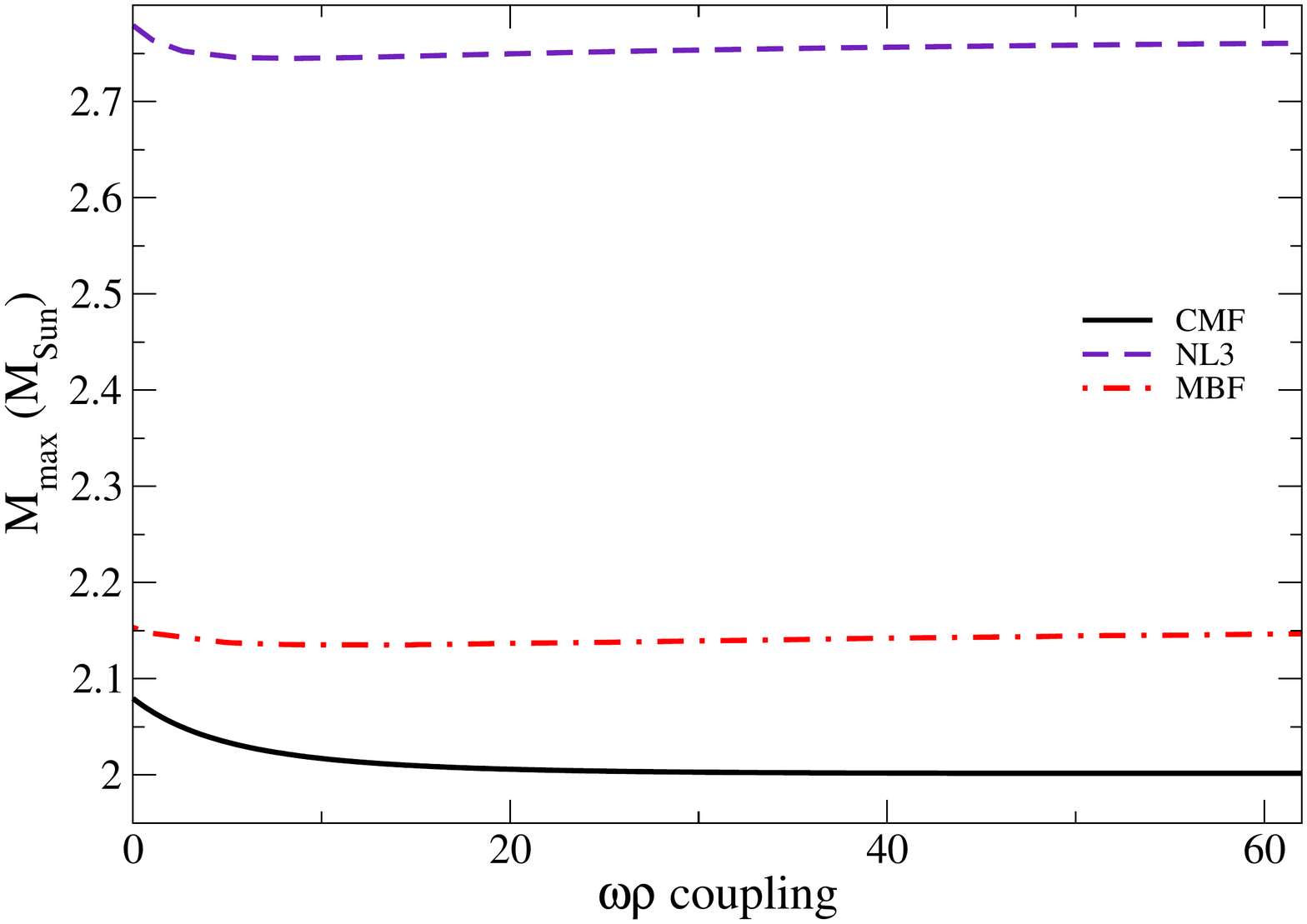} 
%\end{subfigure}
%
\caption{Radius of a $1.4$ M$_\odot$ star (left panel) and maximum stellar mass of sequence (right panel) as a function of the normalized coupling constant for the $\omega\rho$ self-interaction, $g_{\omega\rho} g_\omega g_\rho^2 $.}
\end{figure}

\section{Results}

In the following, we present results from our three different models for charged-neutral matter in chemical equilibrium. The coupling constant for the $\omega\rho$ self-interaction that will appear in our figures is normalized by the value of the standard $\omega$ coupling in each model, becoming $g_{\omega\rho} g_\omega g_\rho^2$. It varies from zero to a large positive value in each model. In the CMF model, for each $g_{\omega\rho}$ coupling, the $g_\rho$ coupling is refitted to keep the same symmetry energy value at saturation (when $g_{\omega\rho}=0$). The $\delta$ meson coupling $g_\delta$ stays the same given by the SU(3) scheme. In the NL3 family, for each $g_{\omega\rho}$ coupling, the $g_\rho$ coupling is refitted by keeping the same symmetry energy value at $n_B=0.1$ fm$^{-3}$ (again, when $g_{\omega\rho}=0$). In the MBF model, for each $g_{\omega\rho}$ coupling, the $g_\rho$ and $g_\delta$ coupling are refitted to keep the same symmetry energy value at $n_B=0.1$ fm$^{-3}$ and at saturation. Other couplings are kept the same in all models.

Macroscopic stellar properties are determined making use of the Tolman-Oppenheimer-Volkoff (TOV) equations and include a crust. For the CMF and MBF models, we added a Baym-Pethick-Sutherland outer crust, together with a Pandharipande inner crust, and a Feynman-Metropolis-Teller atmosphere from Ref.~\cite{Baym:1971pw}. For the NL3 family, the outer crust EoS is again the Baym-Pethick-Sutherland (BPS) \cite{Baym:1971pw} but now we perform a Thomas-Fermi pasta calculation for the inner crust \cite{Grill,Pais:2016xiu}. In order to build the full EoS, the outer crust merges with the inner crust at the neutron drip line and the inner crust merges with the core at the density when the pasta geometries melt, the crust-core transition.

In Fig.~1, we show how the $\omega\rho$ self-interaction strength decreases significantly the radius of a low mass star (by $\sim 7\%$), but at the same time only moderately decreases the maximum stellar mass of the sequence (by $\sim 2\%$). The values of both quantities decrease before approximately saturating for large coupling strengths. This is because a larger $\omega\rho$ coupling strength implies a lower absolute value of the $\rho$ mesonic fields necessary to fit isospin-asymmetric properties around saturation, causing a softening of the EoS's around that region. The maximum mass values saturate for lower coupling strengths than the stellar radius. This is due to the fact that the large $\omega\rho$ coupling has a strong effect at lower densities. Beyond that, it only modifies significantly the EoS at larger densities until a certain coupling strength, after which it has a slight opposite behavior.  Since large-mass stars depend heavily on dense matter EoS, they are not as affected, although for very large couplings the stars become a bit more massive.

\begin{figure}[t!]
\centering
%\begin{subfigure}{0.495\textwidth}
%\centering
\includegraphics[width=0.496\textwidth]{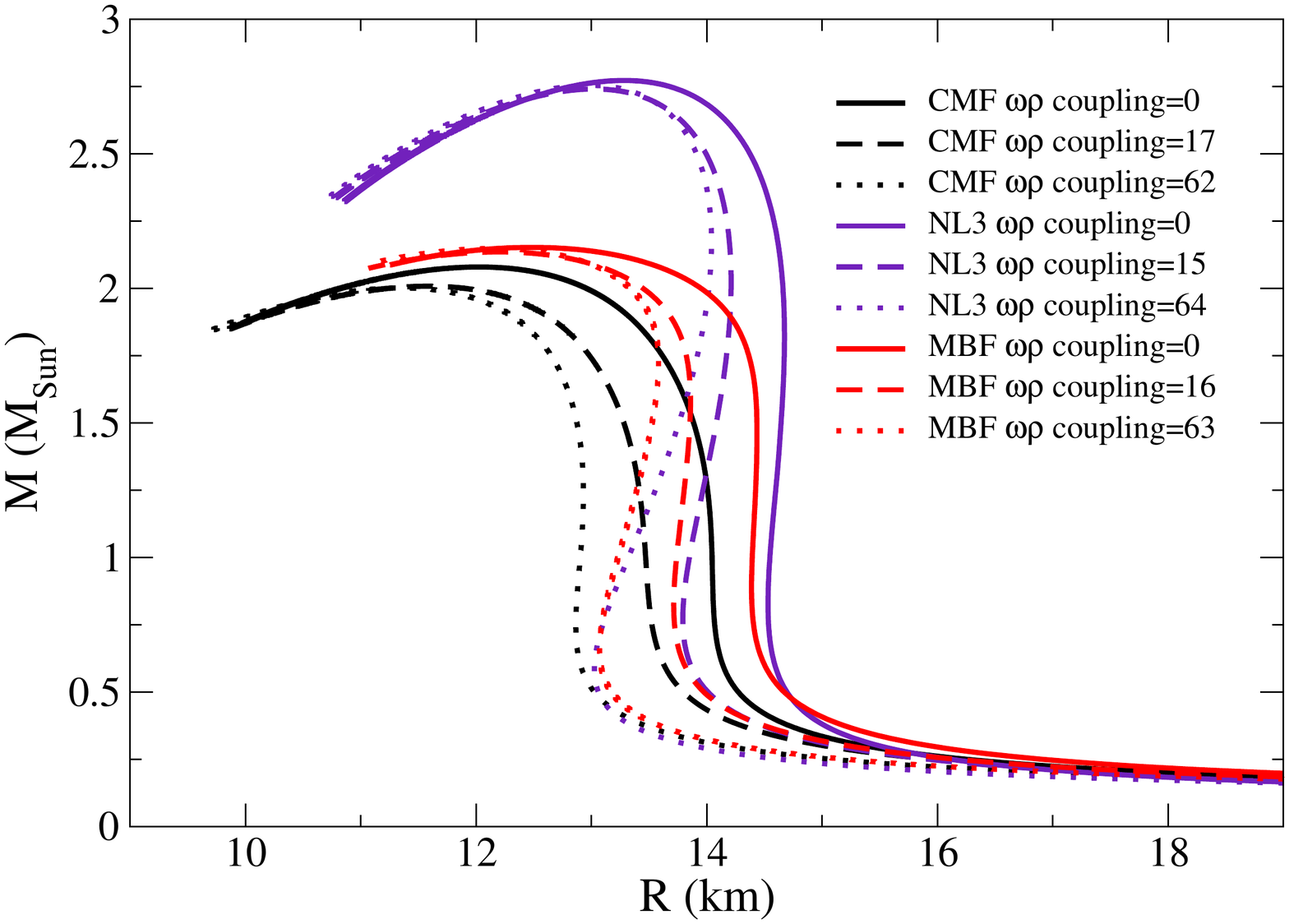} 
%\end{subfigure}
%
%\begin{subfigure}{0.495\textwidth}
%\centering
\includegraphics[width=0.496\textwidth]{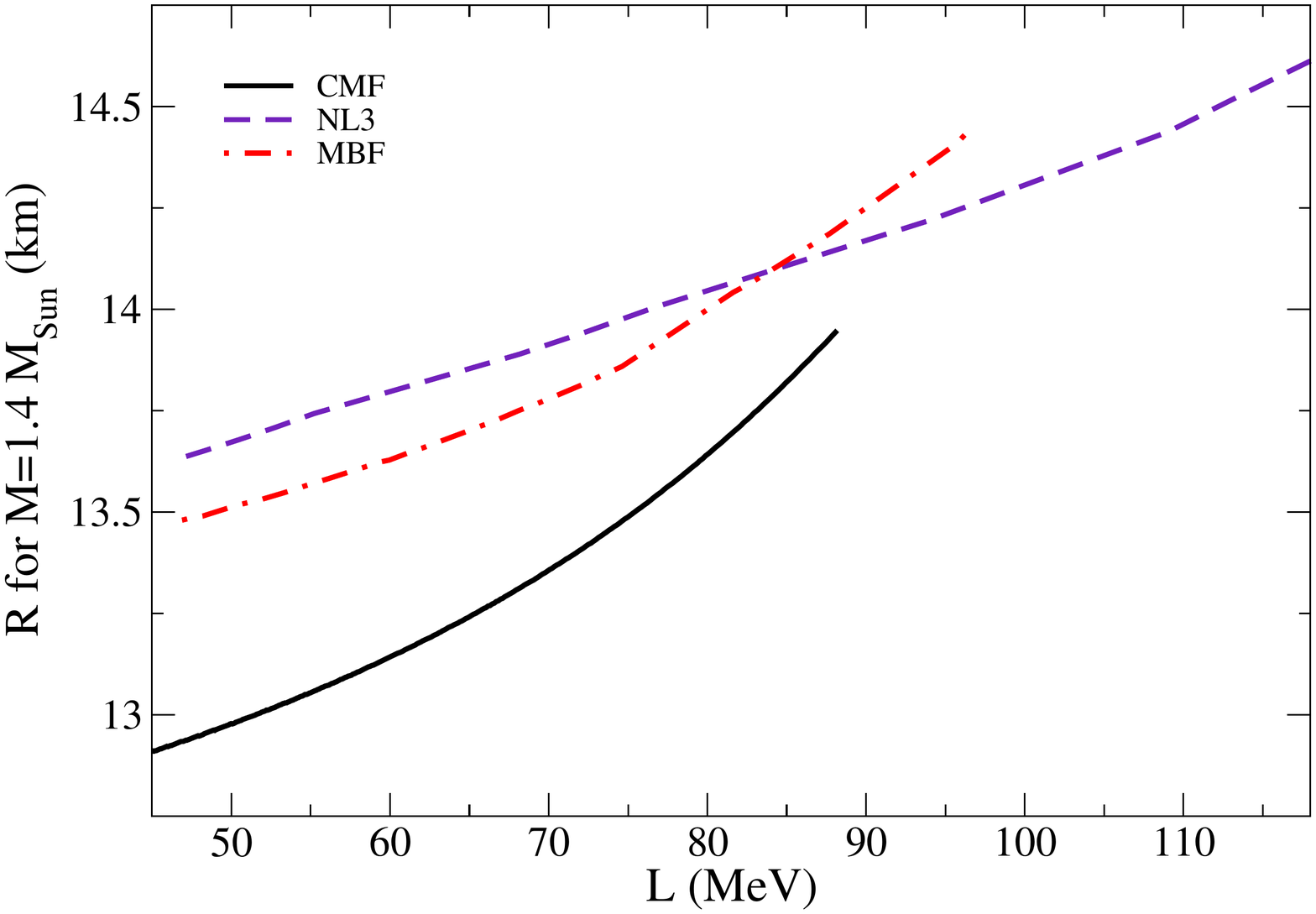} 
%\end{subfigure}
%
\caption{Families of stellar solutions (left panel) and radius of a $1.4$ M$_\odot$ star as a function of symmetry energy slope at saturation density (right panel) for different normalized coupling constants for the $\omega\rho$ self-interaction. In the right panel, the $\omega\rho$ coupling strength runs though the values presented in the horizontal axes of Fig.~1.}
\end{figure}

The relation between  $\omega\rho$ coupling, stellar masses and radii is once more illustrated in the left panel of Fig.~2, where we show stellar families for different $\omega\rho$ self-interaction strengths: zero, an intermediate, and a large value. Note that, for all coupling values analyzed, all models still reproduce $2$ M$_\odot$ stars. The right panel of Fig.~2 shows the relation between radius of a low mass star and slope of the symmetry energy at saturation $L_0=3 n_0 (\partial E_{sym}/\partial n_B)_{n_B=n_0}$. The leftmost points show the slope for the largest $\omega\rho$ coupling analyzed (shown on the x-axis of Fig.~1) and the rightmost points show the slope without the coupling. This relation between stellar radius and symmetry energy slope has already been pointed out by many works, as for example in Refs.~\cite{Steiner:2010fz,Dutra:2014qga,Lopes:2014wda,Dexheimer:2015qha,Wei:2015aep,Alam:2016cli,Margueron:2017lup,Zhang:2018vbw}. Note that Ref.~\cite{Hornick:2018kfi} has recently shown that the nucleon effective mass at saturation has a larger influence (than $L$) in the radius of stars, nevertheless, we cannot verify that, as for our analysis the effective mass is fixed in each of our EoS models.

\begin{figure}[t!]
\centering
%\begin{subfigure}{0.495\textwidth}
%\centering
\includegraphics[width=0.496\textwidth]{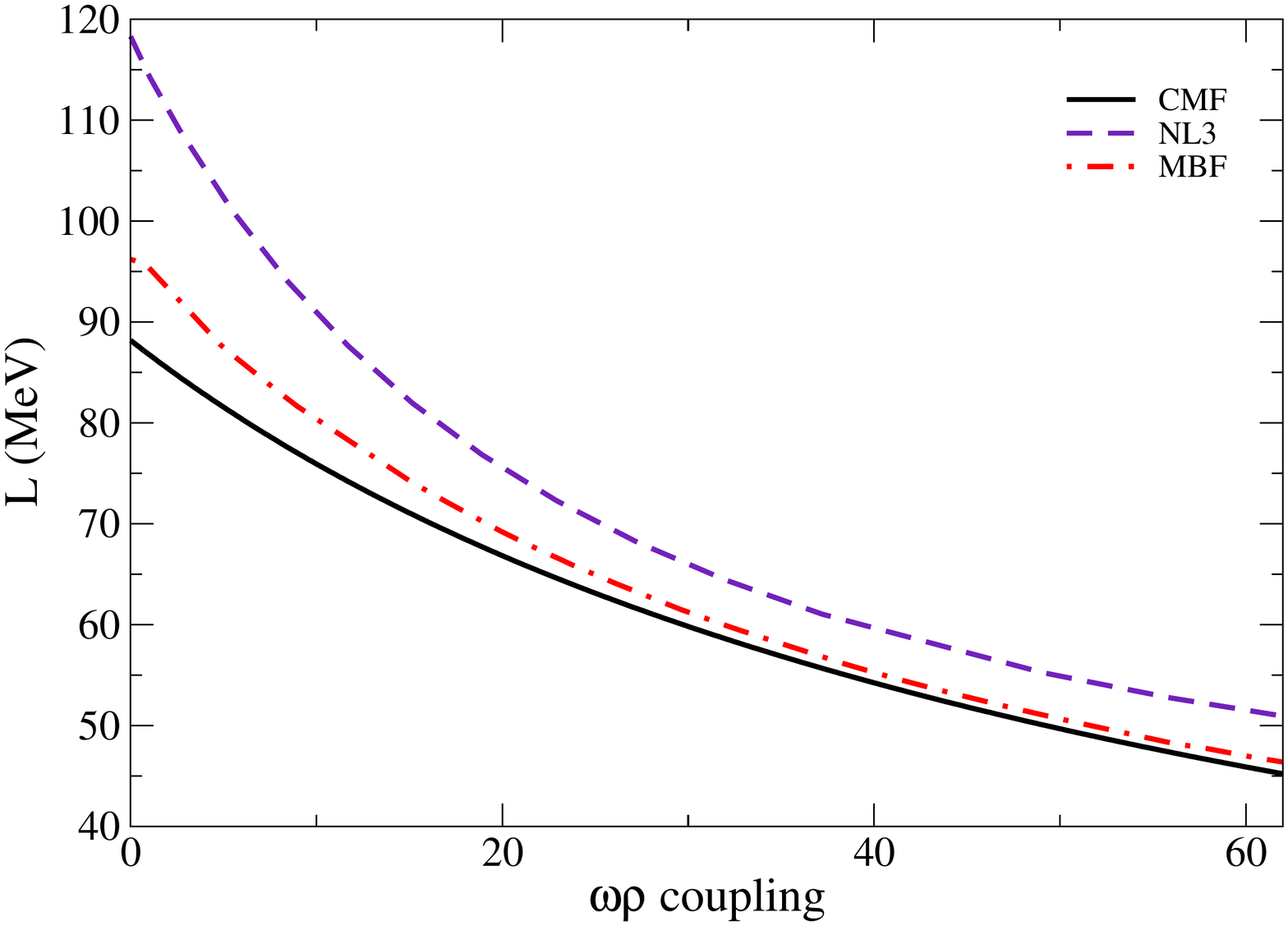} 
%\end{subfigure}
%
%\begin{subfigure}{0.495\textwidth}
%\centering
\includegraphics[trim={0 -1.1cm 0 0},clip,width=0.443\textwidth]{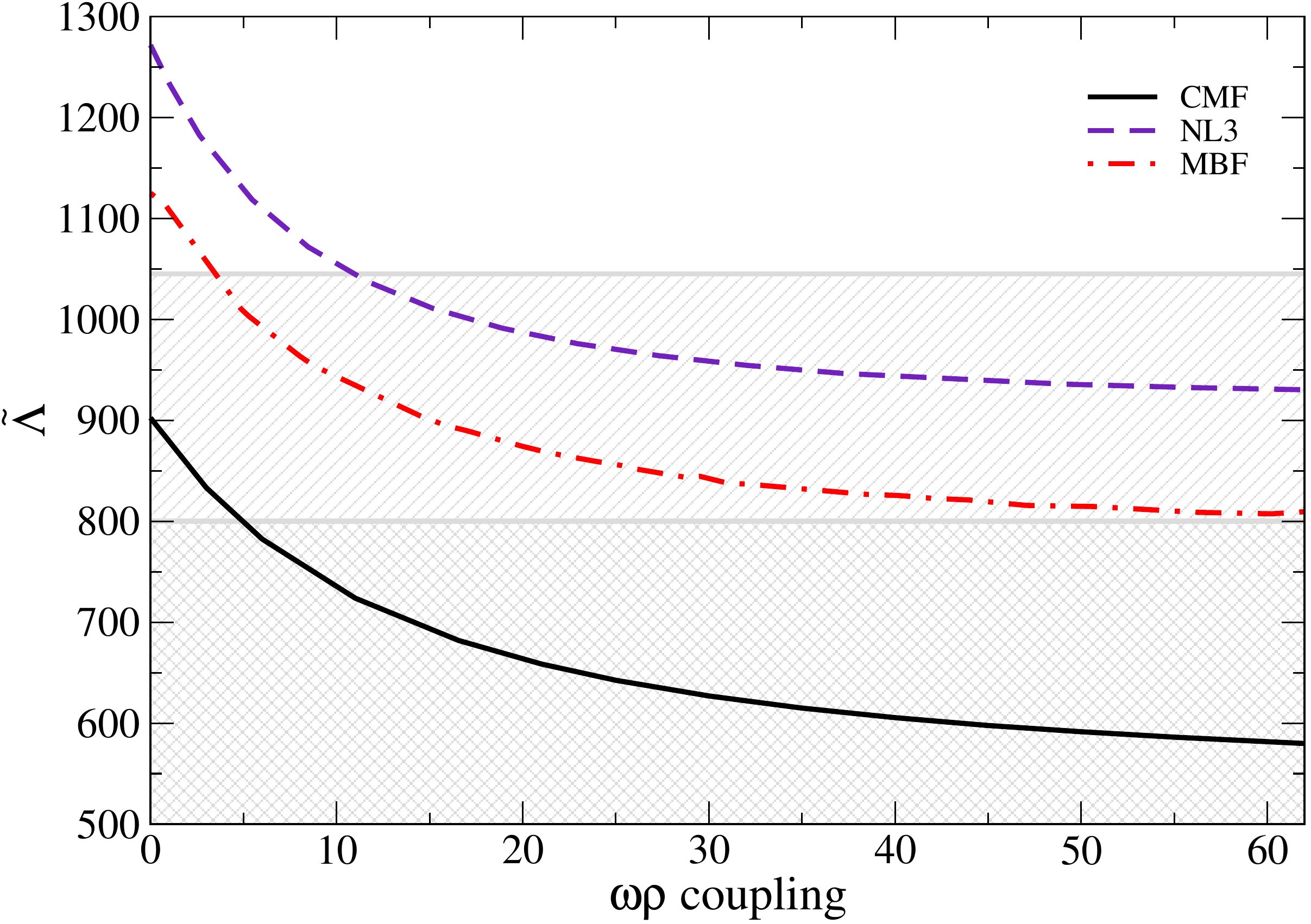} 
%\end{subfigure}
%
\caption{Symmetry energy slope at saturation density (left panel) and tidal deformability of a $1.4$ M$_\odot$ star (right panel) as a function of the normalized coupling constant for the $\omega\rho$ self-interaction. The horizontal grey lines mark the top of the $\tilde{\Lambda}$ range given by the analysis performed in Ref.~\cite{De:2018uhw} and provided in the original LIGO/Virgo paper Ref.~\cite{TheLIGOScientific:2017qsa}.}
\end{figure}

In Fig.~3, we show how the $\omega\rho$ self-interaction strength decreases significantly the slope of the symmetry energy at saturation (by $\sim 53\%$ on average) and also the tidal deformability of a $1.4$ M$_\odot$ star (by $\sim 30\%$) for values of coupling shown in Fig.~1. The dimensionless tidal deformability is calculated using the following expression \cite{Favata:2013rwa}
\begin{equation}
\tilde{\Lambda} = \frac{16}{13}\frac{\left(M_1 + 12 M_2\right)M_1^4 \Lambda_1 + \left(M_2 + 12 M_1\right)M_2^4 \Lambda_2}{\left(M_1 + M_2\right)^5},
\label{eq:Lambda}
\end{equation}
which depends on the mass and dimensionless tidal deformability $\Lambda_i=\lambda_i/M_i^5$ of each star. Both quantities shown in Fig.~3 decrease as a function of the coupling, as they only depend on the matter EoS at low densities. In the right panel, the top horizontal grey line marks the top of the tidal deformability range of $1045$ from Ref.~\cite{De:2018uhw}, discussed in the Introduction. This limit constrains the $\omega\rho$ self-interaction strength to be above $11$ in the NL3 model and above $3$ in the MBF model, but imposes no constraint in the CMF model. On the other hand, a tidal deformability upper bound of $800$, as suggested in the original LIGO/Virgo paper Ref.~\cite{TheLIGOScientific:2017qsa} under the condition of low spin priors (bottom grey line in the figure), constrains the $\omega\rho$ self-interaction strength to be about $70$ in the MBF model and above about $5$ in the CMF model, but is not compatible with the NL3 model.

In addition, Fig.~4  presents how the tidal deformability of a $1.4$ M$_\odot$ star increases with the slope of the symmetry energy and radius of a $1.4$ M$_\odot$ star. Once more, in both panels the rightmost point of each EoS shows the tidal deformability value when the $\omega\rho$ coupling is zero and the leftmost point shows the slope for the largest coupling analyzed (showed in Figs.~1,~3). Although not universal, these relations are extremely useful, because they show that different models respond similarly when the vector/isovector baryon interactions are modified. This allows us to understand better how baryons interact at low and, to some extent, also high densities using current and future measurements of tidal deformability. Some other works along this line investigated relations between tidal deformability and Skyrme interactions \cite{Tsang:2018kqj}, between tidal deformability and specific/symmetry energy coefficients \cite{Zhang:2018vrx,Zhang:2018vbw,Malik:2018zcf}, between the tidal deformability and strangeness \cite{Gomes:2018eiv}, and between tidal deformability and strong interaction strength \cite{Tews:2018iwm}.

\begin{figure}
\centering
%\begin{subfigure}{0.495\textwidth}
%\centering
\includegraphics[trim={0cm -1.2cm 0cm 0},clip,width=0.43\textwidth]{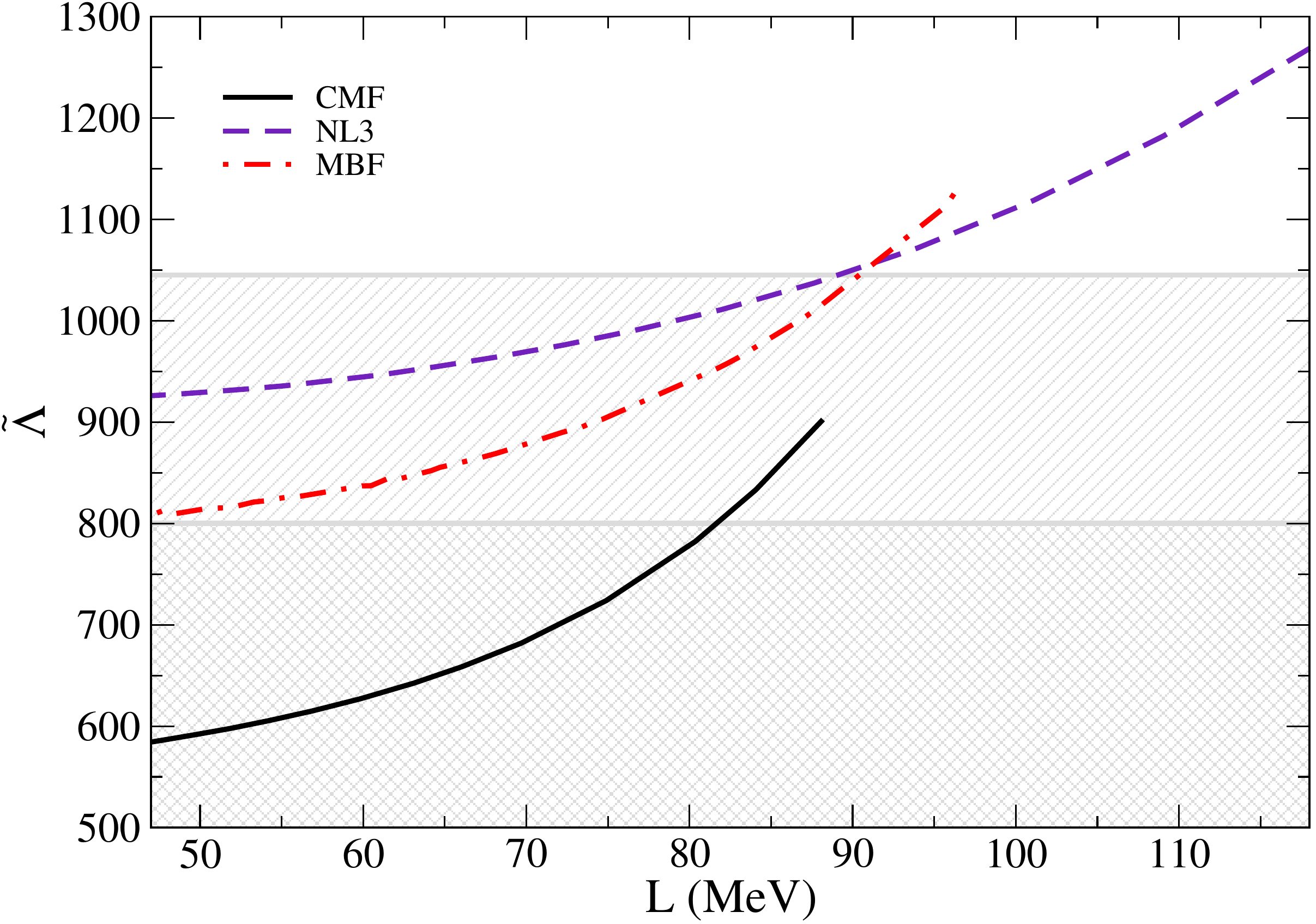} 
%\end{subfigure}
%
%\begin{subfigure}{0.495\textwidth}
%\centering
\hspace{7mm}
\includegraphics[trim={0cm -1.1cm 0 0},clip,width=0.43\textwidth]{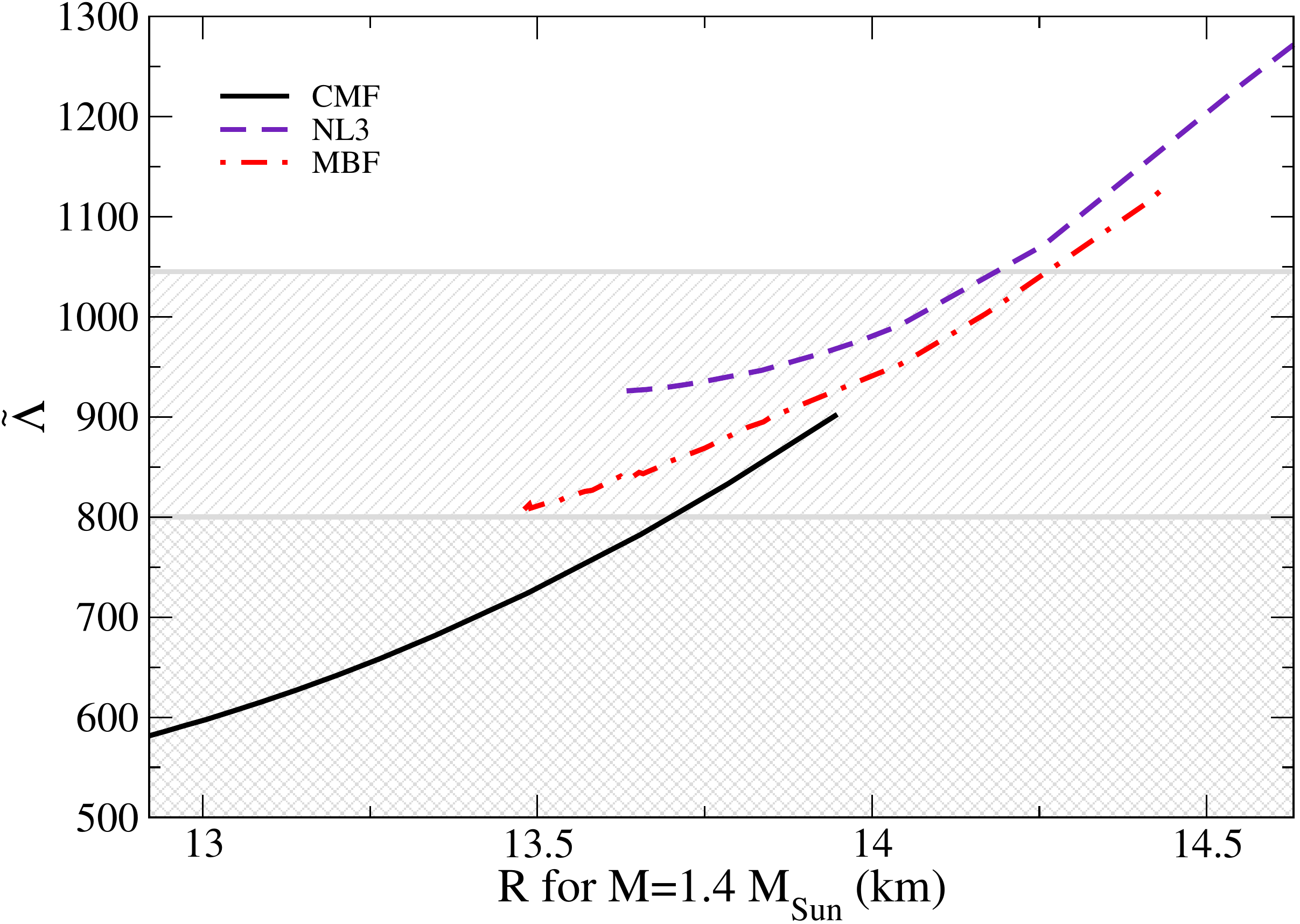} 
%\end{subfigure}
%
\caption{Tidal deformability of a $1.4$ M$_\odot$ star as a function of symmetry energy slope at saturation density (left panel) and as a function of the radius of a $1.4$ M$_\odot$ star (right panel) for different normalized coupling constants for the $\omega\rho$ self-interaction. In both panels the $\omega\rho$ coupling strength runs though the values presented in the horizontal axes of Figs.~1 and 3. Once more, the horizontal grey lines mark the top of the $\tilde{\Lambda}$ range given by the analysis performed in Ref.~\cite{De:2018uhw} and provided in the original LIGO/Virgo paper Ref.~\cite{TheLIGOScientific:2017qsa}.}
\end{figure}

Concerning the horizontal grey lines in Fig.~4, the top of the tidal deformability range of $1045$ constrains the slope of the symmetry energy to be below $90$ MeV in the MBF model and below $89$ MeV in the NL3 model, but imposes no constraint in the CMF model. On the other hand, a tidal deformability upper bound of $800$ constrains the slope of the symmetry energy to be about $43$ MeV in the MBF model and below  $82$ MeV in the CMF model. The top of the tidal deformability range of $1045$ constrains the radius of a $1.4$ M$_\odot$ star to be below $14.3$ km in the MBF and $14.2$ km in the NL3 model, but imposes no constraint in the CMF model. On the other hand, a tidal deformability upper bound of $800$ constrains the the radius of a $1.4$ M$_\odot$ star to be below about $13.5$ km in the MBF model and below  $13.7$ km in the CMF model

As already discussed in the introduction, observations and experiments point to relatively small stellar radii, symmetry energy slope and tidal deformability, all of which decrease as a consequence of increasing the $\omega\rho$ self-interaction strength. In addition, all models with $\omega\rho$ self-interaction are in better agreement with microscopic (chiral effective field theory) neutron matter calculations \cite{Hebeler:2013nza}. This behavior is shown in the left panel of Fig.~5 and stands for all densities provided. Note that instead of trying to compare our results with the results from Ref.~\cite{Hebeler:2013nza} order by order, we have compared our mean-field results with their complete results {(using NN and 3N forces going up to second order)}. In the future, we intend to improve our approach by adding Hartree-Fock corrections to our formalism in a similar fashion as done in Ref.~\cite{Sun:2009jj}.

Finally, the right panel of Fig.~5 shows that the threshold for the direct Urca process increases with the $\omega\rho$ self-interaction. This is important, as an early Urca threshold can trigger this very efficient cooling process in a large portion of intermediate and large mass stars, reproducing stars that are too cold in comparison with observational data \cite{Page:2004fy,Negreiros:2011ak}. Previous works, for example the ones in Refs.~\cite{Newton:2013zaa,Negreiros:2018cho}, have already shown that there is a relation between stellar cooling and the slope of the symmetry energy at saturation.

\section{Conclusions}

In this work, we made use of three different models that take into account different degrees of freedom (nucleons only or the whole baryon octet) and also different features of dense matter, such as a density dependence of coupling constants and chiral symmetry restoration. These models can also reproduce different nuclear, astrophysical, and QCD constraints. We used them to study the impact of a non-linear isovector singlet to isovector triplet coupling of the vector mesons, effectively introducing a density-dependent modification of the isospin sector of the baryonic interactions.

\begin{figure}[t!]
\centering
%\begin{subfigure}{0.495\textwidth}
%\centering
\includegraphics[width=0.496\textwidth]{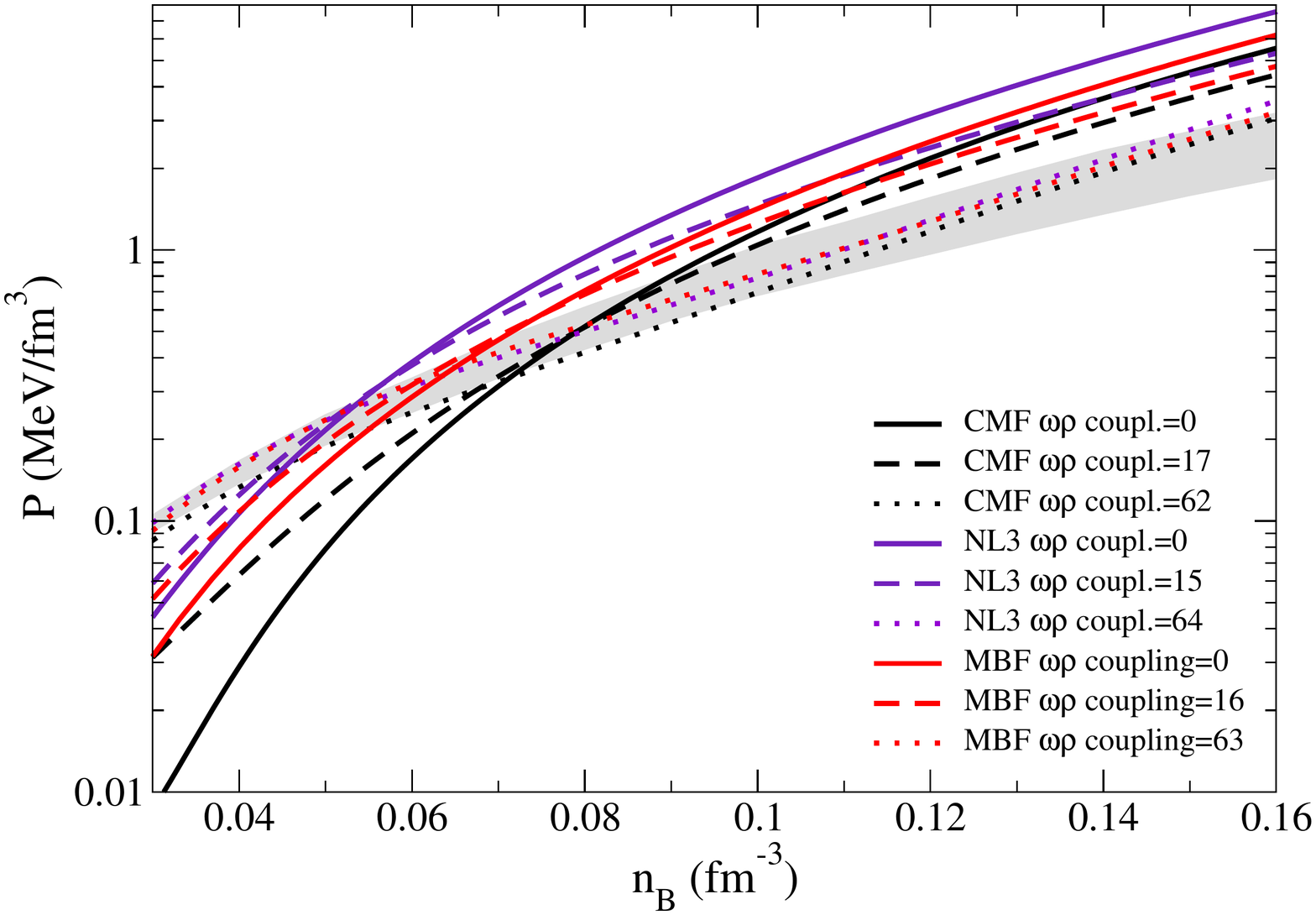} 
%\end{subfigure}
%
%\begin{subfigure}{0.495\textwidth}
%\centering
\includegraphics[width=0.496\textwidth]{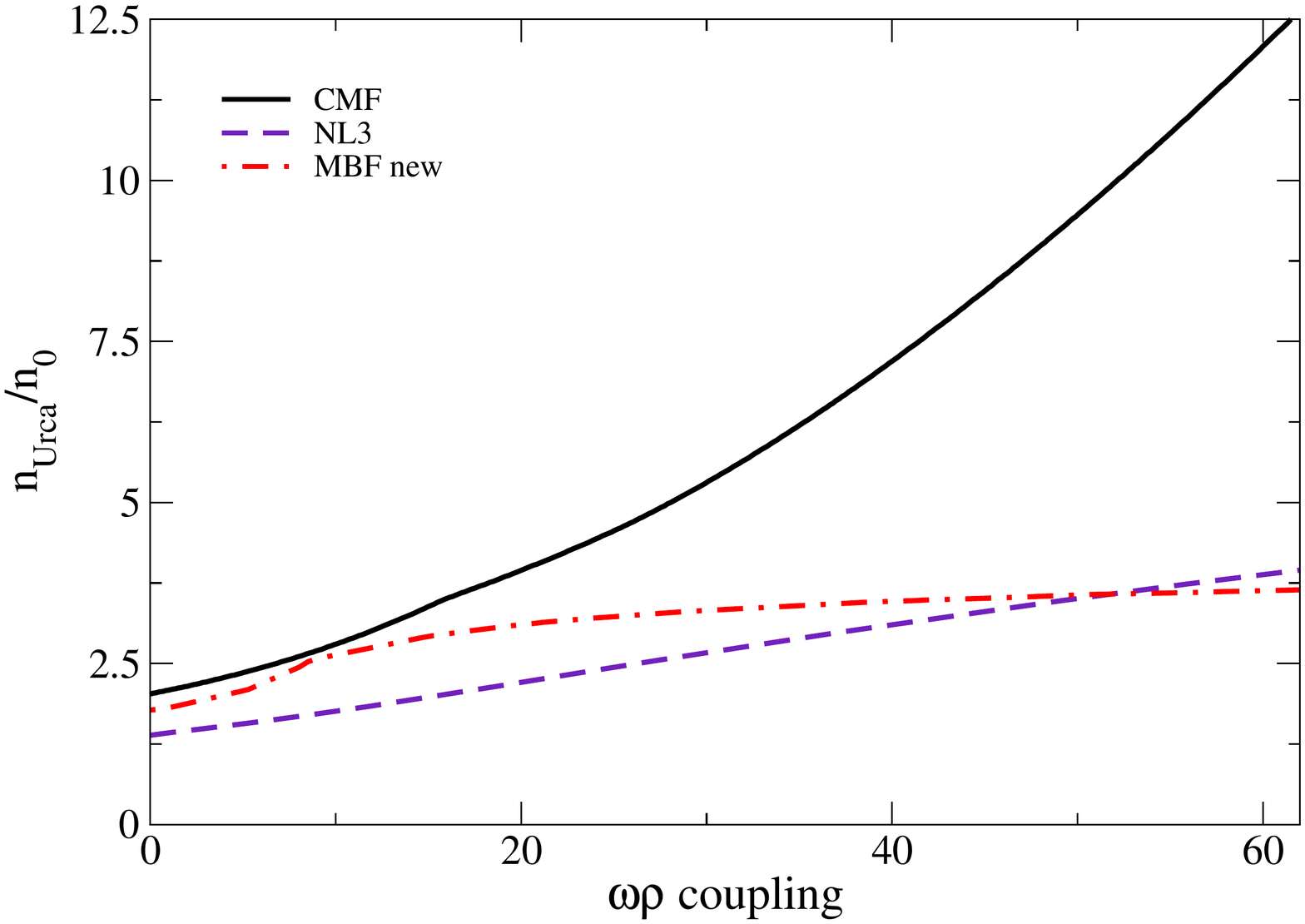} 
%\end{subfigure}

%
\caption{Pressure as a function of baryon number density for pure neutron matter (left panel) and threshold for the direct Urca process in units of saturation density (right panel) for different normalized coupling constants for the $\omega\rho$ self-interaction. In the left panel, limits from Ref.~\cite{Hebeler:2013nza} are additionally shown as a grey region.}
\end{figure}

Overall, independently of the model analyzed, the additional term reduced the radii of neutron stars, in particular that of the ``standard" 1.4 solar mass star, without significantly changing the maximum stellar mass. This then translated to a smaller value of the slope parameter $L$. Similarly, the non-linear coupling led to a reduction of the tidal deformability calculated from a neutron star merger event. We found that values for these three quantities decrease for all models analyzed when the $\omega\rho$ coupling is taken into account, resulting in a better agreement with recent experimental and observational tendencies. Other quantities we analyzed, such as the pure neutron matter equation of state and threshold for direct Urca process, are also in better agreement with chiral effective field theory calculations and stellar cooling data when the $\omega\rho$ coupling is taken into account.

In other words, different constraints on the tidal deformability of neutron stars can translate into $\omega\rho$ coupling strengths and, as a consequence symmetry energy slope and stellar radii values, all of which are model depend but, nevertheless, relevant. This is because, whenever we detect more gravitational waves from neutron star binaries and are able to constrain better the tidal deformability of low and maybe larger mass stars, we will be able to in a more reliable fashion determine which models are in agreement with observations. At the same time, we will be able to learn more through these models about the equation of state for neutron stars at low and high densities, even accounting for rotational effects \cite{Sieniawska:2018pev}). The same will happen when we obtain more accurate neutron star radius measurements from NICER.

\section*{Acknowledgements}

Support comes from PHAROS COST Action CA16214, by the National Science Foundation under grant PHY1748621 {(V. D.)}, by Funda\c c\~ao para a Ci\^encia e Tecnologia (FCT-Portugal) under Project No. SFRH/BPD/95566/2013 {(H. P.)}, and  by the LOEWE center HIC for FAIR.

\section*{References}
\bibliographystyle{iopart-num}
\bibliography{paper}

\end{document}